\documentclass[aps,preprint]{revtex4}
\usepackage{graphicx}
\usepackage{amsmath}
\usepackage{epsfig}
\usepackage{amsfonts}
\usepackage{amssymb}

\begin{document}

\title{Green's functions on finite lattices and their connection to the
infinite lattice limit}
\author{S. Cojocaru}
\affiliation{National Institute of Physics and Nuclear Engineering, Bucharest-Magurele,
Romania}
\affiliation{Institute of Applied Physics, Chi\c{s}in\u{a}u, Moldova}

\begin{abstract}
It is shown that the Green's function on a finite lattice in arbitrary space
dimension can be obtained from that of an infinite lattice by means of
translation operator. Explicit examples are given for one- and
two-dimensional lattices.
\end{abstract}

\maketitle

\section{Introduction}

Lattice Green's functions (LGF) are widely used in solving discrete (e.g.,
Schrodinger) equations. In the infinite lattice (or thermodynamic) limit
explicit expressions for the LGF in different space dimensions $d$ are known
for a long time \cite{Maradudin, Katsura, wortis, mattis}. The form of the
LGF in the direct lattice space ( $X$ ) as obtained by the Fourier transform
method is usually given by a $d-$fold integral over the reciprocal space ($Q$%
) that can be further reduced to known functions. This allows, in
particular, to obtain the eigenenergy equation in an explicit form too (see,
e.g. \cite{wortis} for an example covering $d=1,2,3$ of the Heisenberg spin
model). An extension of this approach to finite lattices has been proposed
in \cite{c1}. It is based on an approximation shown to account sufficiently
well even for subtle details of the exact solution, known from the Bethe
ansatz in the case of a finite spin chain. Control over finite $N$ (the
number of lattice sites along one coordinate) corrections turns to be
essential for a proper description of the wave function symmetry and
behavior of the eigenenergies. The approach has revealed new classes of
excitation modes for lattices in higher dimensions and has allowed to find
their connection to those known from Bethe solutions in $1D$. The further
development of the approach was based on the proof of several exact
expressions for the finite lattice sums and the discrete Fourier transform
in one dimension \cite{c2}. We have recently become aware of the paper \cite%
{ali} where some similar exact formulas have been derived and note that our
paper in \cite{c2} was submitted before publication of \cite{ali}.

In particular, with the discrete Fourier transform (DFT) we reproduce the
exact eigenenergy equations known from Bethe ansatz. The latter, however,
can not be applied to higher dimensional lattices. In the present paper we
generalize the earlier exact formulas to higher dimensions. It turns out
that the GF at finite $N$ can be obtained from that of the infinite lattice
under the action of a simple function of the translation operator. This
result is tested on the $d=1$ case studied earlier and an example for the $%
d=2$ case is also given. Remarkably, the expression for the finite LGF in
two dimensions is obtained in a form of a single integral, as the function
does not reduce to known ones. This form is also efficient in obtaining
exact asymptotic expansions in $N$. The approach is rather general and can
be applied to different lattice types without limitation on space dimension.

\section{Green's function on a finite square lattice}

The form of the LGF studied below%
\begin{align}
G_{sq}\left( X,Y\right) \equiv & \frac{1}{N}\sum_{m=0}^{N-1}\frac{1}{N}%
\sum_{n=0}^{N-1}\frac{\exp\left( iQ_{x}X+iQ_{y}Y\right) }{\lambda
-\cos\left( Q_{x}\right) -\cos\left( Q_{y}\right) },  \label{1} \\
Q_{x} & =\frac{\Delta_{x}}{N}+2\pi\frac{m}{N};\ \ \ Q_{y}=\frac{\Delta_{y}}{N%
}+2\pi\frac{n}{N};  \notag \\
X,Y & =0,1,...,N-1.  \notag
\end{align}
is the simplest one and, at the same time, contains the important $\Delta-$
parameters related to the symmetry of the wave function, as explained in
earlier works \cite{c1,c2}. For simplicity we also confine the energy
parameter $\lambda$ to the region outside the scattering states, $\lambda
\geq2.$ Let's consider a simple generalization of the formula expressing (%
\ref{1}) in terms of modified Bessel functions \cite{c2}
\begin{equation}
G_{sq}\left( X,Y\right) =\int_{0}^{\infty}dze^{-\lambda z}\sum_{q=-\infty
}^{\infty}\sum_{p=-\infty}^{\infty}I_{pN+X}\left( z\right) I_{qN+Y}\left(
z\right) \exp\left( -ip\Delta_{x}-iq\Delta_{y}\right) ,   \label{2}
\end{equation}
that can be trivially extended to higher dimensions. One clearly sees that
in the thermodynamic limit $N\rightarrow\infty$ ( $G_{sq}^{TDL}\left(
X,Y\right) $ ) only terms with $p=0$ and $q=0$ \textquotedblleft survive".
Eq. (\ref{2}) then suggests a possibility to act with the ``translation
operators '' like exp$\left( pN\frac{d}{dX}\right) $ upon the $%
G_{sq}^{TDL}\left( X,Y\right) $ to obtain the full expression at finite $N.$

It should be taken into account that $X$ and $Y$ are defined as above and
that due to its symmetry property the modified Bessel function $I_{n}\left(
z\right) $ should actually be written as $I_{\left\vert n\right\vert }\left(
z\right) $. In this form the action of translation operator on the series is
more transparent. Then by splitting the series in two parts we find from (%
\ref{2}) that%
\begin{equation*}
G_{sq}\left( X,Y\right) =\int_{0}^{\infty}dze^{-\lambda
z}\sum_{q=0}^{\infty}\sum_{p=0}^{\infty}\exp\left( p\left( -i\Delta_{x}+N%
\frac{d}{dX}\right) \right) \exp\left( q\left( -i\Delta_{y}+N\frac{d}{dY}%
\right) \right) I_{\left\vert X\right\vert }\left( z\right) I_{\left\vert
Y\right\vert }\left( z\right)
\end{equation*}%
\begin{equation*}
+\int_{0}^{\infty}dze^{-\lambda
z}\sum_{q=0}^{\infty}\sum_{p=1}^{\infty}\exp\left( -p\left( -i\Delta_{x}+N%
\frac{d}{dX}\right) \right) \exp\left( q\left( -i\Delta_{y}+N\frac{d}{dY}%
\right) \right) I_{\left\vert X\right\vert }\left( z\right) I_{\left\vert
Y\right\vert }\left( z\right)
\end{equation*}%
\begin{equation*}
+\int_{0}^{\infty}dze^{-\lambda
z}\sum_{p=0}^{\infty}\sum_{q=1}^{\infty}\exp\left( p\left( -i\Delta_{x}+N%
\frac{d}{dX}\right) \right) \exp\left( -q\left( -i\Delta_{y}+N\frac{d}{dY}%
\right) \right) I_{\left\vert X\right\vert }\left( z\right) I_{\left\vert
Y\right\vert }\left( z\right)
\end{equation*}%
\begin{equation*}
+\int_{0}^{\infty}dze^{-\lambda
z}\sum_{p=1}^{\infty}\sum_{q=1}^{\infty}\exp\left( -p\left( -i\Delta_{x}+N%
\frac{d}{dX}\right) \right) \exp\left( -q\left( -i\Delta_{y}+N\frac{d}{dY}%
\right) \right) I_{\left\vert X\right\vert }\left( z\right) I_{\left\vert
Y\right\vert }\left( z\right) .
\end{equation*}
Taking the geometric series of translation operators, we obtain%
\begin{equation*}
G_{sq}\left( X,Y\right) =\left[ \frac{1}{1-\exp\left( -i\Delta_{x}+N\frac{d}{%
dX}\right) }+\frac{\exp\left( i\Delta_{x}-N\frac{d}{dX}\right) }{%
1-\exp\left( i\Delta_{x}-N\frac{d}{dX}\right) }\right]
\end{equation*}%
\begin{equation}
\times\left[ \frac{1}{1-\exp\left( -i\Delta_{y}+N\frac{d}{dY}\right) }+\frac{%
\exp\left( i\Delta_{y}-N\frac{d}{dY}\right) }{1-\exp\left( i\Delta_{y}-N%
\frac{d}{dY}\right) }\right] G_{sq}^{TDL}\left( \left\vert X\right\vert
,\left\vert Y\right\vert \right) .   \label{3}
\end{equation}
Note that the operators in the square brackets do not annihilate each other
since they act differently on $G_{sq}^{TDL}\left( \left\vert X\right\vert
,\left\vert Y\right\vert \right) .$ Expressions (\ref{1}) and (\ref{3}) mean
that the LGF at finite $N$ is a linear superposition of the limit LGF
translated with multiples of the lattice period. Generalization of Eq. (\ref%
{3}) to a lattice in an arbitrary dimension is obvious. With this general
representation one can straightforwardly obtain the LGF from the integral
representation of LGF in the thermodynamic limit, e.g.%
\begin{equation}
G_{sq}^{TDL}\left( \left\vert X\right\vert ,\left\vert Y\right\vert \right)
=\int_{0}^{2\pi}\int_{0}^{2\pi}\frac{\exp\left( ip\left\vert X\right\vert
\right) \exp\left( iq\left\vert Y\right\vert \right) }{\lambda-\cos p-\cos q}%
\frac{dp}{2\pi}\frac{dq}{2\pi},   \label{2D}
\end{equation}
using the property
\begin{equation*}
f\left( N\frac{d}{dX}\right) \exp\left( CX\right) =f\left( NC\right)
\exp\left( CX\right) ,
\end{equation*}
and that $\left\vert X-N\right\vert =N-X$ (where $f$ is a function and $C$ a
parameter). One can apply it directly to the known expressions of the $%
G^{TDL}$ for different lattices.

In the $d=1$ case we have
\begin{equation}
G_{1d}^{TDL}\left( \left\vert X\right\vert \right) =\int_{0}^{2\pi }\frac{%
\exp \left( iq\left\vert X\right\vert \right) }{\cosh v-\cos q}\frac{dq}{%
2\pi }=\frac{\exp \left( -v\left\vert X\right\vert \right) }{\sinh v},
\label{4}
\end{equation}%
where the energy parameter $\lambda $ has been replaced by $\cosh v$. Note
that the last equality holds also when the modulus sign on the l.h.s. is
omitted. Thus, one finds%
\begin{equation*}
G_{1d}\left( X\right) =\left[ \frac{1}{1-\exp \left( -i\Delta +N\frac{d}{dX}%
\right) }+\frac{\exp \left( i\Delta -N\frac{d}{dX}\right) }{1-\exp \left(
i\Delta -N\frac{d}{dX}\right) }\right] \frac{\exp \left( -v\left\vert
X\right\vert \right) }{\sinh v}
\end{equation*}%
\begin{equation}
=\left[ \frac{\exp \left( -vX\right) }{1-\exp \left( -i\Delta -Nv\right) }+%
\frac{\exp \left( i\Delta +\left( X-N\right) v\right) }{1-\exp \left(
i\Delta -Nv\right) }\right] \frac{1}{\sinh v},  \label{5}
\end{equation}%
i.e. the result obtained earlier \cite{c2}. For the sake of comparison to
higher dimensional lattice we also give the result of applying the operator
to the first equality in (\ref{5}) and leading to the same result.%
\begin{equation*}
\int_{0}^{2\pi }\frac{\exp \left( iqX\right) }{\cosh v-\cos q}\times \frac{1%
}{1-\exp \left( -i\Delta +iqN\right) }\frac{dq}{2\pi }
\end{equation*}%
\begin{equation*}
+\int_{0}^{2\pi }\frac{\exp \left( -iqX\right) }{\cosh v-\cos q}\times \frac{%
\exp \left( i\Delta +iqN\right) }{1-\exp \left( i\Delta +iqN\right) }\frac{dq%
}{2\pi }.
\end{equation*}

However in the case of square lattice such a direct application of
translation operators to $G_{sq}^{TDL}\left( \left\vert X\right\vert
,\left\vert Y\right\vert \right) $ does not reduce to a known function.
Generally, $G_{2D}^{TDL}$ is a double hypergeometric Appel function \cite%
{Katsura}, that can be reduced to elliptic integrals in more simple cases
(as the one considered in the present paper). Then we can use the approach
defined by (\ref{3}) and discussed below. An equivalent way would be to
substitute in (\ref{2}) the well known integral representations of Bessel
function involving exponentials \cite{W}%
\begin{equation*}
I_{n}\left( z\right) =\frac{1}{2\pi }\int_{0}^{2\pi }\exp \left( z\cos
y\right) \exp \left( izy\right) dy,
\end{equation*}%
\begin{equation*}
I_{n}\left( at\right) I_{m}\left( at\right) =\int_{0}^{2\pi }\frac{dy}{2\pi }%
I_{n+m}\left( 2at\cos y\right) \exp \left( -i\left( n-m\right) y\right) ,
\end{equation*}%
to finally reduce $G_{sq}\left( X,Y\right) $ to a single integral.

In analogy with the $d=1$ case, according to (\ref{3}) and (\ref{2D}), we
obtain ($\gamma =2/\lambda $)%
\begin{equation*}
G_{sq}\left( X,Y\right) =\frac{1}{N}\sum_{m=0}^{N-1}\frac{1}{N}%
\sum_{n=0}^{N-1}\frac{\exp \left( iQ_{x}X+iQ_{y}Y\right) }{\lambda -\cos
\left( Q_{x}\right) -\cos \left( Q_{y}\right) }=
\end{equation*}%
\begin{equation}
\frac{\gamma }{4\pi }\int_{0}^{2\pi }\frac{d\phi }{\sqrt{1-\left( \gamma
\cos \phi \right) ^{2}}}\times   \label{main}
\end{equation}%
\begin{align*}
& \left( \frac{\exp \left( Xf^{\ -}\right) }{\left( 1-\exp \left( -i\Delta
_{x}+Nf^{\ -}\right) \right) }+\frac{\exp \left( i\Delta _{x}+\left(
N-X\right) f^{\ -}\right) }{\left( 1-\exp \left( i\Delta _{x}+Nf^{\
-}\right) \right) }\right)  \\
& \times \left( \frac{\exp \left( Yf^{\ +}\right) }{\left( 1-\exp \left(
-i\Delta _{y}+Nf^{\ +}\right) \right) }+\frac{\exp \left( i\Delta
_{y}+\left( N-Y\right) f^{\ +}\right) }{\left( 1-\exp \left( i\Delta
_{y}+Nf^{\ +}\right) \right) }\right) ,
\end{align*}%
where
\begin{equation*}
f\left( \phi \right) \equiv \ln \left( \frac{\gamma \cos \phi }{1+\sqrt{%
1-\left( \gamma \cos \phi \right) ^{2}}}\right) ,
\end{equation*}%
\begin{equation*}
f^{\ +}\equiv f\left( \phi \right) +i\phi ;\ \ \ f^{\ -}\equiv f\left( \phi
\right) -i\phi .
\end{equation*}%
The structure of this exact expression is surprisingly simple and indicates
the possibility of extension to other lattices. In particular, (\ref{main})
contains the periodic symmetry of the LGF represented by the transformation $%
X\rightarrow \left( N-X\right) $ and $Y\rightarrow \left( N-Y\right) .$ For $%
X,Y<<N$ \ and $\gamma \neq 1\ $we find in the thermodynamic limit, $%
N\rightarrow \infty :$%
\begin{equation}
G_{sq}^{TDL}\left( X,Y\right) =\frac{\gamma }{4\pi }\int_{0}^{2\pi }\frac{%
\exp \left( i\phi \left( Y-X\right) \right) }{\sqrt{1-\left( \gamma \cos
\phi \right) ^{2}}}\left( \frac{\gamma \cos \phi }{1+\sqrt{1-\left( \gamma
\cos \phi \right) ^{2}}}\right) ^{X+Y}d\phi .  \label{6}
\end{equation}%
Eq. (\ref{6}), the "byproduct" of (\ref{main}), gives at the same time a
convenient integral representation for the hypergeometric function $_{4}F_{3}
$ - the explicit form of the $G_{sq}^{TDL}\left( X,Y\right) $ found in \cite%
{Katsura}, that arises in various other contexts, see, e.g. \cite{Barsan}.
In particular, for $X=Y=0$ we find the known result%
\begin{equation*}
G_{sq}^{TDL}\left( 0,0\right) =\frac{2}{\pi \lambda }K\left( \frac{2}{%
\lambda }\right)
\end{equation*}%
Another important observation is that in the $N\rightarrow \infty $ limit
the symmetry related information contained in the $\Delta $ -parameters
seems to be lost. However, recalling our earlier results, the slow $N-$%
convergence of the the "correction" terms in the vicinity of the critical
condition $\gamma =1$ proves that the symmetry of the corresponding
solutions of the Schrodinger equation on a two-dimensional lattice is
retained even in the thermodynamic limit.

In conclusion, the main result of the paper is the derivation of new exact
expressions for the Green's function on a finite two-dimensional lattice. It
is shown that these expressions are the result of applying a simple function
of translation operator to the Green function of the infinite lattice. The
derivation outlines the way for generalizations to other lattice types and
different space dimensions. The exact formulas reduce to known functions and
solutions in particular cases. These expressions allow to obtain asymptotic
expansions. The latter are straightforward away from the critical parameter
region and can be found by standard methods in the vicinity of critical
values of parameters. The integral formulas can be easily adapted to the
region of the energy parameter corresponding to the "scattered states", i.e.
$\lambda <2.$These further developments will be considered in other
publications.

\end{document}